# Local and macroscopic tunneling spectroscopy of $Y_{1-x}Ca_xBa_2Cu_3O_{7-\delta}$ films: evidence for a doping dependent *is* or *id*$_{xy}$ component in the order parameter

A. Sharoni[1], O. Millo[1*]

A. Kohen[2], Y. Dagan[2], R. Beck[2], G. Deutscher[2**]

G. Koren[3]

[1]*Racah Institute of Physics, The Hebrew University, Jerusalem 91904, Israel*

[2]*School of Physics and Astronomy, Raymond and Beverly Sackler Faculty of Exact Science, Tel Aviv University, 69978 Tel Aviv, Israel*

[3]*Department of Physics, Technion - Israel Institute of Technology, Haifa 32000, Israel*

ABSTRACT

Tunneling spectroscopy of epitaxial (110) $Y_{1-x}Ca_xBa_2Cu_3O_{7-\delta}$ films reveals a doping dependent transition from pure $d_{x^2-y^2}$ to $d_{x^2-y^2}+is$ or $d_{x^2-y^2}+id_{xy}$ order parameter. The subdominant (*is* or *id*$_{xy}$) component manifests itself in a splitting of the zero bias conductance peak and the appearance of subgap structures. The splitting is seen in the overdoped samples, increases systematically with doping, and is found to be an inherent property of the overdoped films. It was observed in both local tunnel junctions, using scanning tunneling microscopy (STM), and in macroscopic planar junctions, for films prepared by either RF sputtering or laser ablation. The STM measurements exhibit fairly uniform splitting size in [110] oriented areas on the order of 10 nm$^2$ but vary from area to area, indicating some doping inhomogeneity. U and V-shaped gaps were also observed, with good correspondence to the local faceting, a manifestation of the dominant *d*-wave order parameter.

* e-mail: milode@vms.huji.ac.il

**e-mail: guyde@post.tau.ac.il

## I. Introduction

While it is well established by now that the order parameter (OP) of YBa$_2$Cu$_3$O$_{7-\delta}$ (YBCO) (and other high T$_C$ superconductors) has a dominant $d_{x^2-y^2}$ (hereafter, $d$) component [1-3], the existence and nature of a subdominant OP component is still under debate. The $d$-wave nature of YBCO, as well as the appearance of a subdominant component, should be clearly reflected in directional dependent tunneling spectroscopy, which yield direct information on the quasi-particle density of states [4]. The hallmark of a pure $d$-wave OP in the tunneling spectra is the zero bias conductance peak (ZBCP) observed for tunneling along and near the nodal [110] direction [5-11], reflecting the existence of surface Andreev bound states at the Fermi level [12-16]. In addition, U and V-shaped gaps can be seen, respectively, in the [100] and [001] directions [8, 11, 15, 17, 18].

Deviations from a pure $d$-wave behavior were found experimentally and predicted theoretically by various groups. $is$ or $id_{xy}$ (hereafter, $id`$) components, proposed by Fogelstroem [19] and Laughlin [20], respectively, both remove the nodes of the dominant $d$-wave OP and manifest themselves by splitting the ZBCP. On the other hand, a subdominant $s$-wave OP that was also proposed [3] does not remove (only shifts) the nodes, and therefore has only little effect on the ZBCP. Spontaneous splitting of the ZBCP (in zero magnetic field) was observed by a few groups [5, 11, 21-24]. This splitting was observed only below some critical temperature ~10 K, and was attributed to a transition to a state of broken time reversal symmetry with $d+is$ or $d+id`$ OP. However, other groups did not find any splitting of the ZBCP [6-8, 10]. Deutscher and collaborators have shown that such splitting takes place only beyond some critical doping level, close to optimal [21, 24, 25], and the dependence of the split on the magnetic field (both magnitude and direction) is better explained by a $d+id`$ OP [9, 24, 25]. Sharoni et al. also noted that ZBCP splitting is not observed for underdoped YBCO films [11]. Evidence for a $d+s$ OP in the overdoped regime was claimed by Yeh et al., based on $c$-axis tunneling data. However, tunneling along the nodal [110] direction was not reported for overdoped samples in their work [26].

We have performed an extensive tunneling study of optimally doped and Ca-overdoped (110) YBCO films prepared using two different deposition techniques, RF sputtering and laser ablation, and two types of tunnel junctions, macroscopic (planar) and local (using STM). We find, irrespective of tunneling junction type or film deposition method, a clear correlation between the hole doping and the spontaneous splitting of the ZBCP, with the splitting increasing linearly with doping in the overdoped regime. Furthermore, the values of the splits obtained by the two distinct tunneling methods are in good agreement with each other. These data confirm those of Dagan and Deutscher obtained on oxygen-overdoped (110) YBCO [24], as well as the preliminary results obtained by Deutscher et al. [25] and Kohen et al. [22] on Ca-doped films. Evidence for a $d+is$ (or $d+id`$) OP is found not only in [110] tunneling, but also in tunneling spectra that we have measured in off-nodal directions. Thus, our data strongly indicate that a $d+is$ (or $d+id`$) OP is an intrinsic property of YBCO in the overdoped regime. Yet, our zero-field tunneling experiments cannot distinguish, by themselves, between the $is$ or $id`$ scenarios.

## II. Experimental

*Film growth*

Epitaxial thin (110) oriented $Y_{1-x}Ca_xBa_2Cu_3O_{7-\delta}$ (Ca-YBCO) films, with different hole doping levels, were grown by either RF sputtering or laser ablation. The hole doping level was controlled by varying the Ca content and for the films prepared by laser ablation also via post-growth oxygen annealing.

The sputtered films were grown on (110) $SrTiO_3$ substrates using off-axis RF sputtering. A $Pr_1Ba_2Cu_3O_{6-\delta}$ template was used in order to reduce misorientations, as described in Ref. 27. The films were examined using x-ray diffraction, which showed only peaks corresponding to the desired (110) orientation. Both optimally doped (x=0) and overdoped films (5% and 10% Ca) were studied. Atomic force microscope (AFM) and scanning electron microscope (SEM) measurements of these films reveal a high density of craters, cracks and nanocracks on the surface, as shown in Fig. 1. Figure 1(a) is a SEM image, 15x15 $\mu m^2$ in size, exhibiting a well defined orientation of these cracks, running in the [001] direction (this was confirmed by directional resistance measurements [25, 27]). One can also observe larger surface defects, but these are well separated from each other (much more than the typical scan range in our STM measurements). A more detailed view of the surface morphology is given by the 1.5x1.5 $\mu m^2$ AFM images of Figs. 1(b) and 1(c), where well oriented nanocracks (b) and craters (c) are observed.

The second type of films were prepared with 30% Ca doping by laser ablation on a (110) $SrTiO_3$ substrate at a temperature of 800 C. The samples were annealed in 0.5 atm. of oxygen and cooled down at a rate of 300 C/h, with a dwell of 15 hours at 420 C before cooling down to room temperature. The surface topography was governed by features similar to those of the sputtered films, as depicted by the AFM image of side 0.3 $\mu m$ in Fig. 1(d).

Figure 2 shows the temperature dependence of the resistance for films with different Ca doping. As one can see, $T_c$ reduces with increased doping, indicating an increase in the hole doping of the samples. Another effect of increasing the Ca doping is a broadening of the transition width, indicating enhanced spatial fluctuations in the hole doping across the sample. In the optimally doped sample $T_c$ is around 90 K and the transition is relatively sharp (~ 2 K), while for the 30% Ca-YBCO sample the transition onset is around 70 K and the width is as high as 15 K.

*Tunneling spectroscopy*

Tunneling spectra (*dI/dV* vs. *V* characteristics) were acquired using either local or macroscopic planar tunnel junctions. In cases where both techniques were applied (on the sputtered films), for the purpose of comparison, two films were prepared in the same run, thus having identical properties.

The macroscopic tunnel junctions were prepared as follows. Immediately after the film growth, we pressed an indium pad against the cuprate film, with an approximate contact area of 1 $mm^2$. This process results in junctions with typical resistance ranging from a few Ohms to a few tens of Ohms, yielding reproducible tunneling spectra. The junctions are stable on the scale of a few weeks and can undergo a number of thermal cycles with no significant change in tunneling characteristics. I(V) curves were measured digitally using a current source, and were differentiated numerically to obtain the *dI/dV - V* spectra. Each measurement comprised of two successive cycles, to check the absence of heating-hysteresis effects.

The local tunneling measurements were performed using a cryogenic home-made STM, providing spatial resolution of less than 1 nanometer for the tunneling spectra.

The samples were transferred directly from the growth chamber into a dry oxygen ambient in an over-pressured chamber, then mounted within a few hours into our STM and cooled down to 4.2 K via He exchange gas. The tunneling spectra were acquired either directly by the use of conventional lock-in technique, or by numerical differentiation of the measured I-V curves, with similar results obtained by both methods. We have confirmed that the measured gaps and ZBCP features were independent of the STM voltage and current setting (before disconnecting momentarily the feedback circuit). This rules out the possibility that the gap features are due to the tunneling condition (e.g., tip-sample distance), such as in the case of the Coulomb blockade [28]. STM morphology images were taken before and after acquiring the local tunneling spectra to confirm the exact position of spectra acquisition. All the spectra and topographic images were obtained with a set bias well above the superconductor gap (~25-50 meV), and the (normal) tunneling resistance varied between 100 MΩ to 1GΩ.

## III. Results and discussion

*Morphology and local tunneling spectra correlation*

In a previous work, Sharoni et al. found good correlation between the local tunneling spectra and the surface morphology for (001) YBCO films grown by laser ablation [11]. Here, we demonstrate that such a correlation holds also for (110) Ca-YBCO films prepared by both laser ablation and RF sputtering, focusing on features manifesting *is* (or *id$_{xy}$*) components in the OP. This is presented in Fig. 3 for a Ca-YBCO RF sputtered film and in Fig. 4 for a Ca-YBCO laser ablated film.

Figure 3(a) shows a STM image taken on a sputtered 5% Ca-YBCO sample with (110) nominal surface, focusing on a crack. The crack runs along the [001] direction, meaning that the side-walls expose most likely the (100) plane, while the nanocrack probably ends in a (001) surface. However, the (1n0) or (0n1) planes, respectively, are also possible.

Far from the crack, on the (110) surface, split ZBCPs are typically observed, characteristic of *is* or *id`* subdominant OP. In Fig. 3(b) we plot 10 tunneling spectra (solid lines) measured at locations covering a region of 10x10 nm$^2$ around position 1 in the STM image. The two lower curves (dashed and solid lines) were measured sequentially at the same position, manifesting the reproducibility of our measurements. It is evident that the split ZBCP structure is well reproduced over this area, with the split size, $\delta_s$, varying between 3 to 3.3 meV. However, the structure at higher bias shows significant spatial variations, both in the magnitude (with respect to the ZBCP) and sharpness of the gap-edge structure. Other regions have shown similar behavior, although the average local $\delta_s$ was in some cases smaller. There were also regions, mainly close to the cracks, where splitting was not found, for example the spectra displayed in Fig. 3(c), taken along the crack around position 2 in Fig. 3(a). This may be due to local oxygen deficiency, reducing the local doping beneath the critical value for the onset of a broken time reversal symmetry state. As in Fig. 3(b), the ratio between the ZBCP height to that of the gap-structure varies spatially. This may be due to nano-faceting of the (110) surface [11] or to the effect of local roughness or disorder [19, 29, 30].

Further manifestation of the spectroscopy-morphology correlation is provided by the V-shaped gap structures measured at the edge of the nanocrack (e.g., solid curve in Fig. 3(d) taken at position 3), consistent with *c*-axis tunneling. This is in agreement with the nanocracks running in the [001] direction, see above. There were also measurements taken inside the nanocracks that displayed U-shaped gaps with no in-gap peaks, corresponding to tunneling in the [100] direction (dashed curve), consistent with the wall orientation.

The 30% Ca-doped laser ablated YBCO film also depicted a rich variety of tunneling spectra, showing good correlation with surface morphology and clearly exhibiting a significant contribution of *is* (*id`*) OP. The STM image in Fig. 4(a) shows nanocracks running along the [001] direction, as marked by the arrow. Spectrum 4(b) was measured on the flat (110) surface, at point 1, showing a split ZBCP structure. We note that this split, with peak-to-peak separation $\delta_s$=5.4 meV, is the largest we have measured, in good correspondence with the high doping level (see below).

In addition to the splitting of ZBCP, the *is* (*id`*) subdominant OP can influence also the gap structures, as shown by spectrum 4(c), and in a more subtle way by 4(d), each acquired within and near an edge of a nanocrack in positions 2 and 3, respectively. The dominant feature in spectrum 4(c) is the sharp sub-gap peak structure. This spectrum can be well reproduced (left inset of Fig. 4(c)) using the extended BTK model [14, 15], assuming a *d+is* OP and tunneling into the (120) plane.

Recall that (120) facets may well exist on the side-walls of a nanocrack. The parameters used in this fit are $\Delta_d$ = 17 meV, $\Delta_s$ = 0.14$\Delta_d$ and a small life-time broadening (Dynes) parameter [31], $\Gamma$ = 0.03 $\Delta_d$. We note that equally good fits to our data (for all curves) were obtained assuming either an *is* or *id`* subdominant OP, so we can not distinguish here between these two options.

A similar sub-gap peak structure was observed by Yeh et al., for tunneling (nominally) in the [001] direction on an overdoped film, and was attributed to a *d+s* OP [26]. We find this interpretation less likely in our case, since in order to fit the positions of the sub-gap peaks (around 1.8 meV), a very large contribution of the subdominant *s* component, more than 0.5$\Delta_d$, is needed (assuming $\Delta_d$ retains its average value observed for this sample, ~ 15 meV). This would also push the shoulders of the outer (main) measured gap much further out as compared to our experimental observation (see right inset of Fig. 4(c)).

Spectrum 4(d) (taken at position 3) exhibits, at first sight, a simple V-shaped gap. However, a closer look reveals a smeared kink structure at low energies (onset around 5 mV), which is not expected for tunneling into a pure *d*-wave superconductor (see, e.g., right inset of Fig. 4(d)). This feature appeared many times in our measurements of overdoped samples, with abundance increasing with doping. This spectrum also conforms well with the picture above, of a *d+is* (*d+id`*) OP and tunneling to a (120) plane, as depicted by the good fit (dotted line), obtained with $\Delta_d$ = 16 meV, $\Delta_s$ = 0.1$\Delta_d$ and $\Gamma$ = 0.15 $\Delta_d$. The main difference between this spectrum and the one presented in Fig. 4(c) is the larger (yet still reasonable) broadening parameter needed to account for the data. This broadening, which may be due to larger local disorder, results in smearing of the sub-gap peak structure. Another possible way to account for the spectrum in 4(d) is by considering c-axis tunneling and, *again*, a *d+is* OP. This is a reasonable scenario, since tunneling may have taken place to the (001) facet at the end of the nanocrack (the tip was positioned a few nm from there). A simulated spectrum for this case, calculated with $\Delta_d$ = 14 meV, $\Delta_s$ = 0.05$\Delta_d$ and $\Gamma$ = 0.05$\Delta_d$, is given in the left inset of Fig. 4(d). Although this curve well reproduces the low-energy kink, it does not fit the experimental data as accurately as the one obtained assuming the (120) tunneling (note, however, that smaller $\Gamma$ and $\Delta_d$ values were used in the fit for the *c*-axis tunneling).

*Doping dependence of ZBCP split*

We now turn to discuss the doping dependence of the subdominant *is* (*id`*) OP, associated with a state of broken time reversal symmetry. The most pronounced evidence for the existence of this subdominant OP is the splitting of the ZBCP in the [110] directional tunneling spectra, where the splitting increases with this subdominant component. We will also compare data obtained from the `macroscopic` planar tunnel junctions (prepared by indium press) with those acquired locally using STM, focusing on the shape and split-size of the ZBCP.

Typical results are presented for (nominally) optimally doped YBCO and 5% Ca-YBCO RF sputtered films in Figs. 5 and 6, respectively. The data for the 10% Ca-doped samples were similar to those of the 5% ones, with typically larger split values. In both figures, spectrum (a) was measured on the `macroscopic` junction, while (b-d) are local (STM) tunneling spectra. (Note again that for each doping level, the films used for both measurements were prepared together in the same run, hence they were nominally identical.) In all measurements performed on the indium pressed macroscopic junction, the ZBCP appears inside a pronounced gap structure. This was also the case in many, but not in all, of our STM spectra. In particular, we could find *dI/dV-V* characteristics that appear very similar to those measured on the corresponding planar junction (see Figs. 5(b) and 6(b)). However, the peak to gap-edge ratio and the general asymmetry of the curves (that vary spatially) usually differ from the `macroscopic` features (e.g., curves (c) and (d) in both figures, as well as Figs. 3(b) and 3(c)). Our STM topographic measurements support this conjecture, exhibiting unit-step size surface roughness on large parts of the films. Note that we have shown previously that even a single unit-cell step affects the local tunneling spectra [11].

The tunneling characteristics of the (nominally) optimally doped films (Fig. 5) showed no splitting of the ZBCP, for both types of tunnel junctions. However, some other (nominally) optimally doped YBCO films prepared by laser ablation [11] did show splitting with small peak-to-peak separations, up to $\delta_s$=1.2 meV. It seems thus that a transition to a state of broken time reversal symmetry takes place (at least at 4.2 K) at a critical doping around optimal, in agreement with previous results [24, 25].

The spectrum in Fig. 6(a), measured on a sputtered 5% Ca-YBCO using planar junction, shows a clear splitting of the ZBCP with $\delta_s$ = 3.2 meV. The STM data exhibit splits with $\delta_s$ ranging between 2 meV to 3.3 meV (6(c) and 6(d) respectively) with many in the vicinity of 3.2 meV, as in Figs. 6(b) and 3(b). These results demonstrate that the local and the macroscopic tunnel junctions yield similar results, thus the splitting of the ZBCP is not a property of the type of tunnel junction. We do find a distribution of $\delta_s$ values in the local tunneling measurements, around the `macroscopic` value, probably due to spatial fluctuations of the doping [32]. The macroscopic tunnel junction thus provides some kind of a mean split size, averaging over the local doping fluctuations to which the STM is sensitive. The doping fluctuations are also reflected in the R(T) measurements (Fig. 2), where the transition width is expected to increase with the magnitude of the fluctuations across the film. Indeed, the $\delta_s$ variation in the 5% Ca-YBCO films (2 - 3.3 meV) is somewhat smaller than that observed for the 10% samples (2.2 - 3.7 meV). Interestingly, the transition for the 10% Ca-YBCO film starts at a temperature where the 5% sample is in the middle of the transition (83 K) and indeed there is an overlap of the $\delta_s$ values measured for these two films. The 30% samples exhibit significantly larger $\delta_s$ variations (3.1 to 5.4 meV), in correspondence with the markedly broader transition width (Fig.2). As already mentioned, this sample was prepared by laser ablation, so we can conclude that the preparation method has no influence on the main features related to time reversal symmetry breaking.

From what we have presented it is clear that $\delta_s$ increases with doping. For a more quantitative picture we use the relation $(p-p_o)^2 \propto ( T_{c,max} - T_c )$, where $p$ is the oxygen content per unit cell, $p_o$ is that of optimal doping, and $T_{c,max}$ is the transition temperature of the optimally doped samples. A plot of the average split size (measured by STM) as a function of $(T_{c,max} - T_c)$, presented in Fig. 7, reveals a clear linear dependence, signifying that $\delta_s$ is proportional to the doping level. In addition to the data obtained for the (110) Ca-YBCO films discussed above (full circles), we also added results measured [11] on (110) facets of (001) grown YBCO films (empty circles). We emphasize that although the data presented in Fig. 7 were measured on films prepared by different deposition methods with different orientations, and the hole doping was controlled by different procedures, all data points fall on the same line. Results shown in Fig. 7 are in agreement with previous results of Dagan and Deutscher, [24] obtained for (110) YBCO films (not Ca doped), and support their interpretation in terms of a quantum critical point for the onset of a broken time reversal symmetry state.

## IV. Conclusions

Our experiments demonstrate the importance of correlating the local tunneling spectra with surface morphology for deciphering the electronic properties of high temperature superconductors, in particular the OP symmetry. Knowing only the nominal growth direction of films may not be enough in order to understand the local spectra measured on them, unless they are perfectly flat, and this is never the case.

The comparison between spatially resolved STM measurements and macroscopic tunneling measurements also proves itself to be highly effective in studies of the above issues. While the former yield important information on the local doping and local morphology dependence of the tunneling characteristics, the latter clearly manifest the `averaged` features associated with the main tunneling direction.

Our data provide clear evidence for the existence of a doping-driven transition from a pure $d$-wave OP to a state with a subdominant $is$ or $id_{xy}$ OP in Ca-YBCO films, associated with a breaking of time reversal symmetry. At 4.2 K, this transition occurs at a critical doping level close to optimal, and the relative contribution of the subdominant component, manifested by the size of the split of the ZBCP, increases linearly with doping at the overdoped side of the YBCO phase diagram. This is in agreement with the conclusions of Ref. 24, drawn from experiments performed on oxygen-doped films. This behavior appears to be independent of the film deposition method (RF sputtering or laser ablation), doping procedure (oxygen treatment or Ca doping), nominal film-growth direction ([110] or [001]), as well as of the type of tunnel junction used (macroscopic indium pressed planar junction or microscopic junctions formed by the STM tip). Therefore, the OP transition described above is an intrinsic property of YBCO, or at least an inherent surface property of this material, and may be due to quantum criticality near optimal doping.

*Acknowledgements*: This work is supported by the Israel Science Foundation, Center for Tunneling Phenomena in Nanostructured Materials and Devices, and the Heinrich Hertz-Minerva Center for High Temperature Superconductivity.

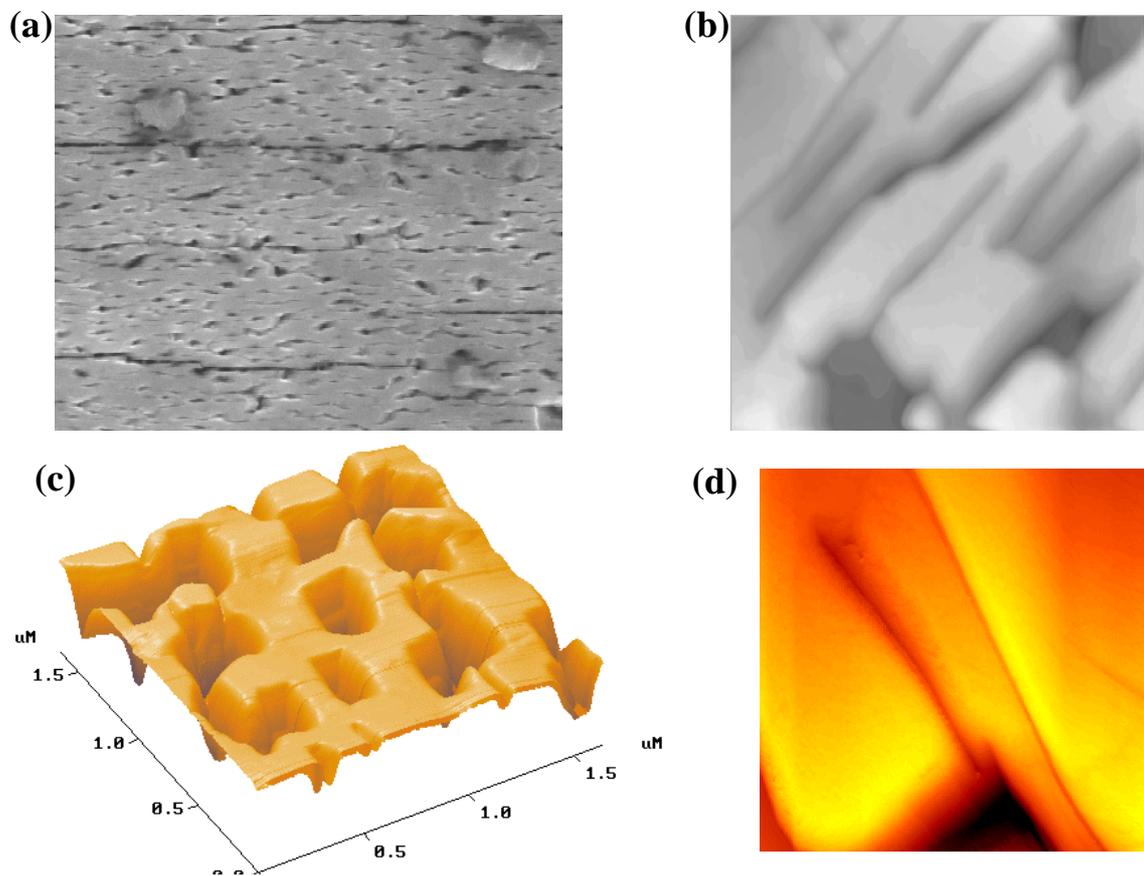

FIG 1. (a) 15x12 µm² TEM image of an RF sputtered film, reveling craters and nanocracks. (b) and (c) 1.5x1.5 µm² AFM images of RF sputtered samples, focusing on nanocracks and craters, respectively. (d) 0.3x0.3 µm² AFM image of a film prepared by laser ablation.
12

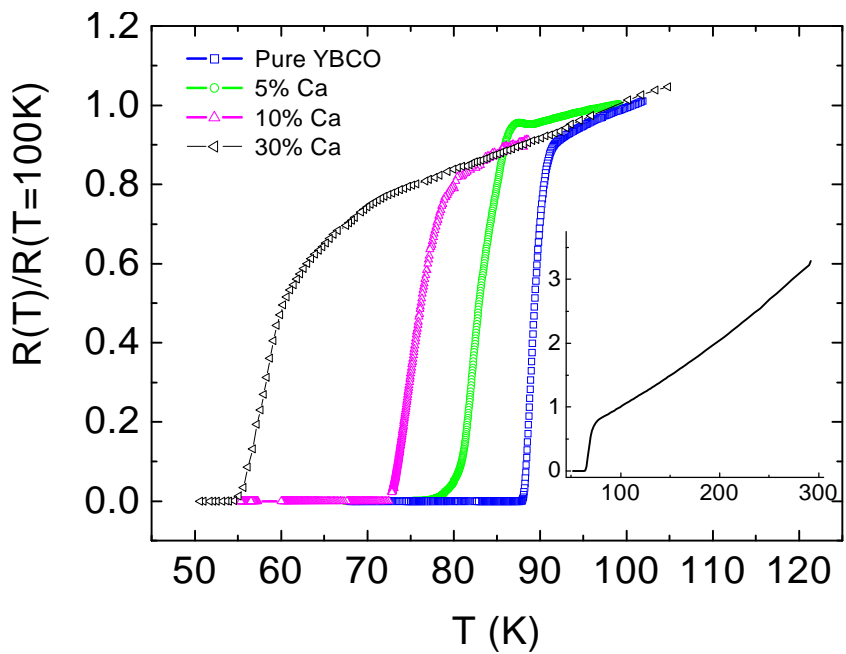

FIG 2. Normalized (to R at T=100 K) transport measurements of optimally doped and Ca doped YBCO films, as denoted in the top left corner of the image. Hole doping increases beyond optimal with Ca content, thus reducing $T_C$. The 30% Ca-doped sample was prepared by laser ablation, whereas the other three were prepared by off-axis RF sputtering. The inset shows a larger temperature range for the R(T) of the 30% Ca-doped sample, with a positive curvature, typical of overdoped films (all overdoped samples exhibited similar behavior).

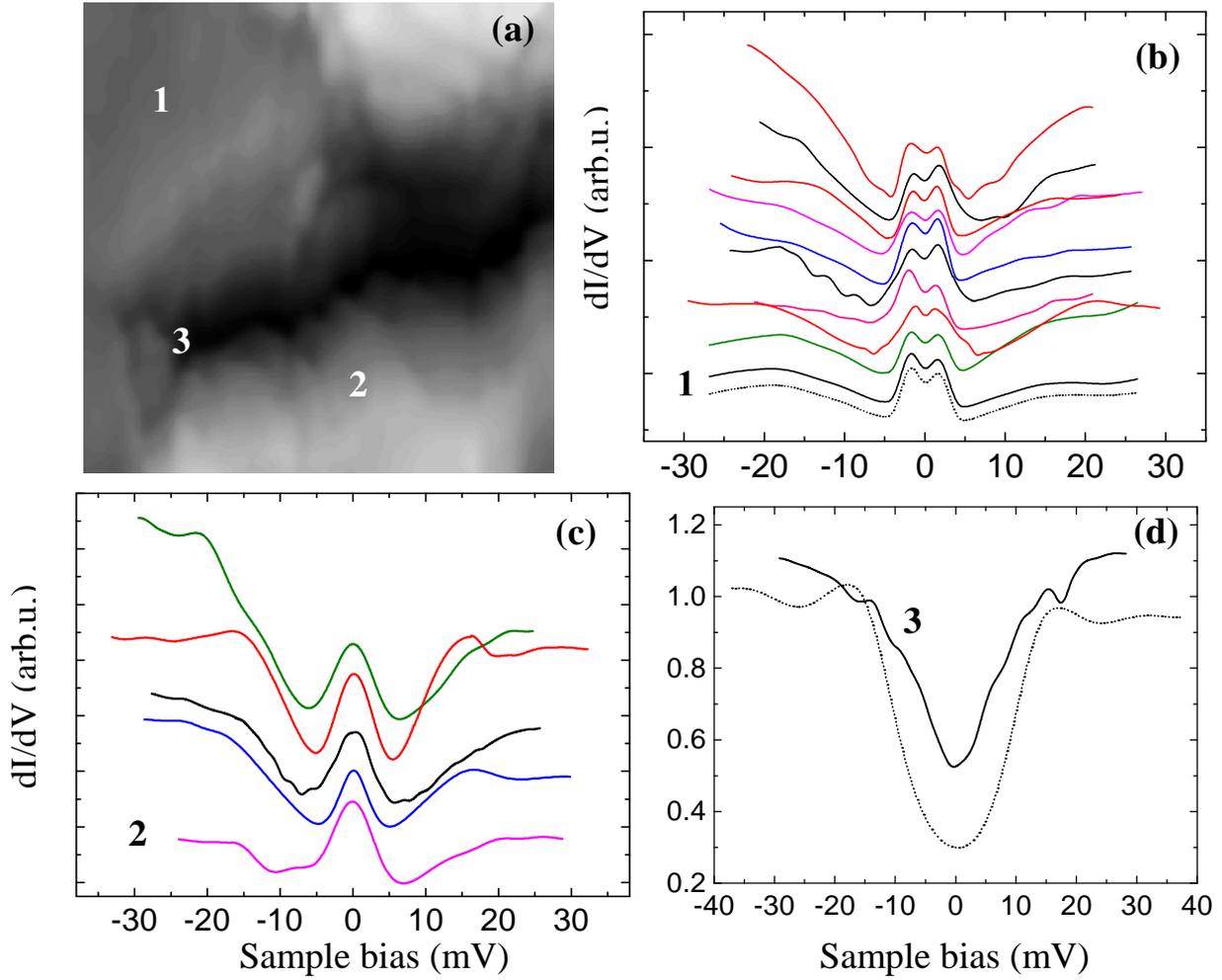

FIG. 3. Correlation between surface morphology of a RF sputtered 5% Ca-YBCO (110) film and the local tunneling spectra. (a) A 0.4x0.4 μm$^2$ STM topographic image focusing on a crack running in the [001] direction. The spectra shown in (b), taken on a 10x10 nm$^2$ area around position 1, exhibit reproducible split ZBCP structure, a manifestation of *d+is* (*d+id`*) OP. The two lower curves (dashed and full) were taken sequentially at the same point, showing the reproducibility of our data. All curves were normalized to the same peak height and shifted vertically for clarity. (c) Spectra taken along the crack-edge, in a range of 10 nm around position 2. (d) A V-shaped gap (full line) measured near the end of the crack (point 3), where a (001) facet is expected to exist. Inside cracks, on (100) side-walls, U-shaped gaps were found, e.g. the doted curve.

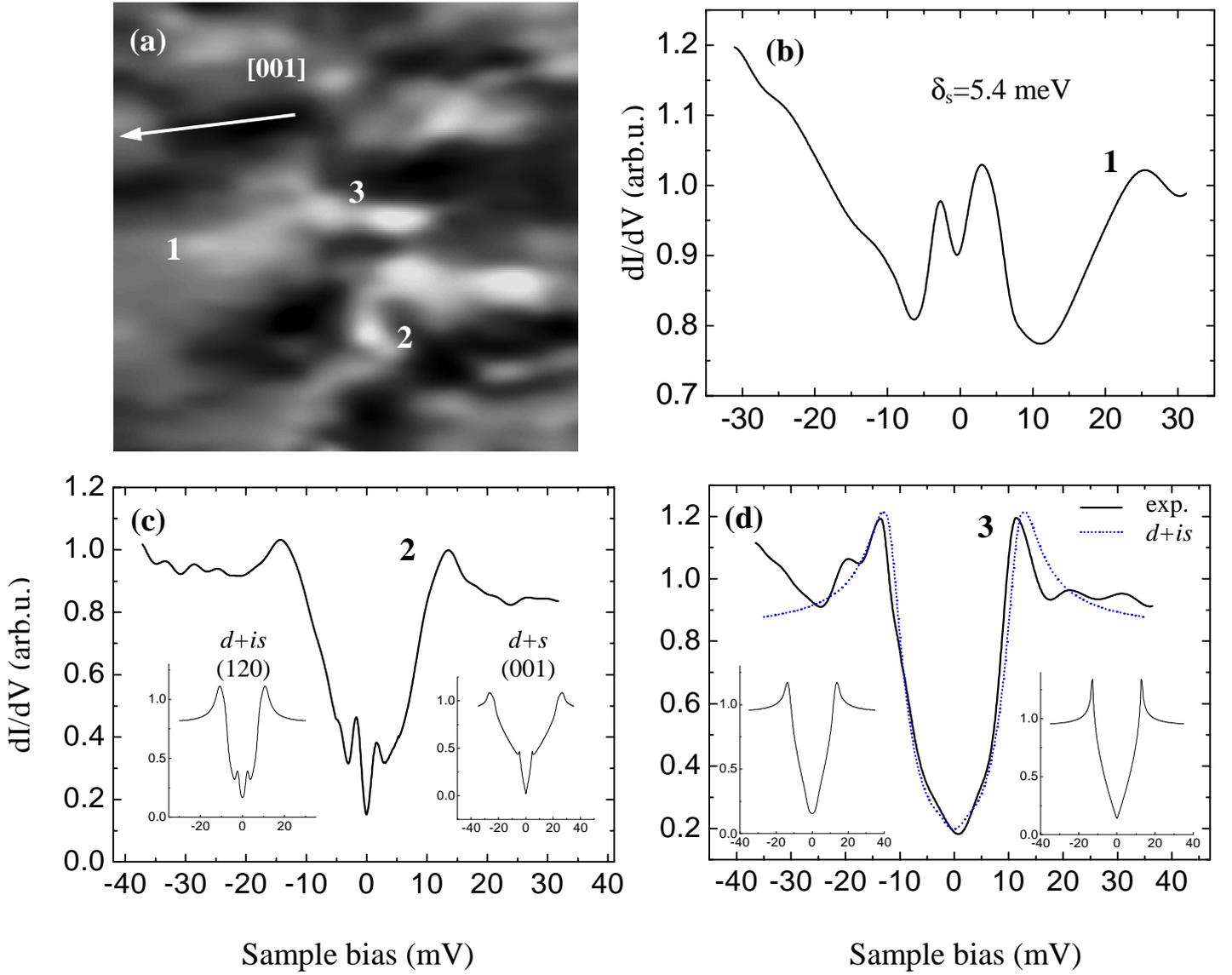

FIG 4. Correlation between surface morphology of 30% Ca-YBCO film prepared by laser ablation and local tunneling spectra. The 120x120 nm$^2$ topography image (a) reveals nanocracks running in the [001] direction, as marked. (b) Spectrum measured at point 1 on the (110) surface, showing the largest ZBCP split we have observed for this sample, with $\delta_s$=5.4 meV. In (c) and (d) we present manifestations of the *is* (*id`*) subdominant OP in directional tunneling spectra other than the [110] (see text). Curve (c) is best explained by tunneling into the (120) plane, as shown in the left inset. The right inset shows, for comparison, simulation for *c*-axis tunneling with a *d+s* OP. Curve (d), taken at position 3, can be fit relatively well to a *d+is* OP assuming either tunneling into the (120) plane (dotted curve) or [001] tunneling (left inset). The right inset shows, for comparison, pure *d*-wave *c*-axis tunneling.

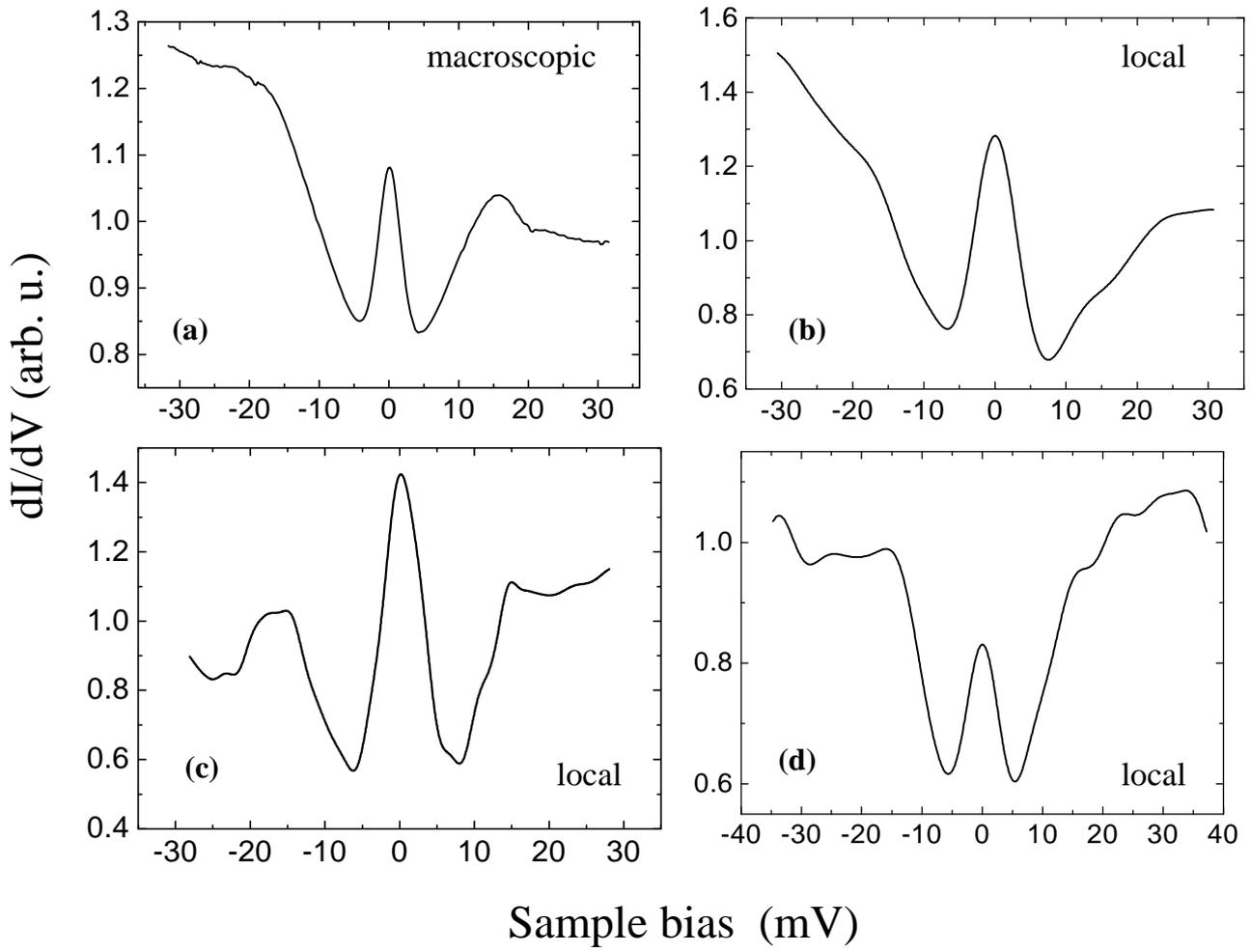

FIG 5. Tunneling spectrum measured on optimally doped RF sputtered YBCO film, using indium pressed planar junction (a) and three spectra acquired with the STM at different locations on a nominally identical film (b-d).

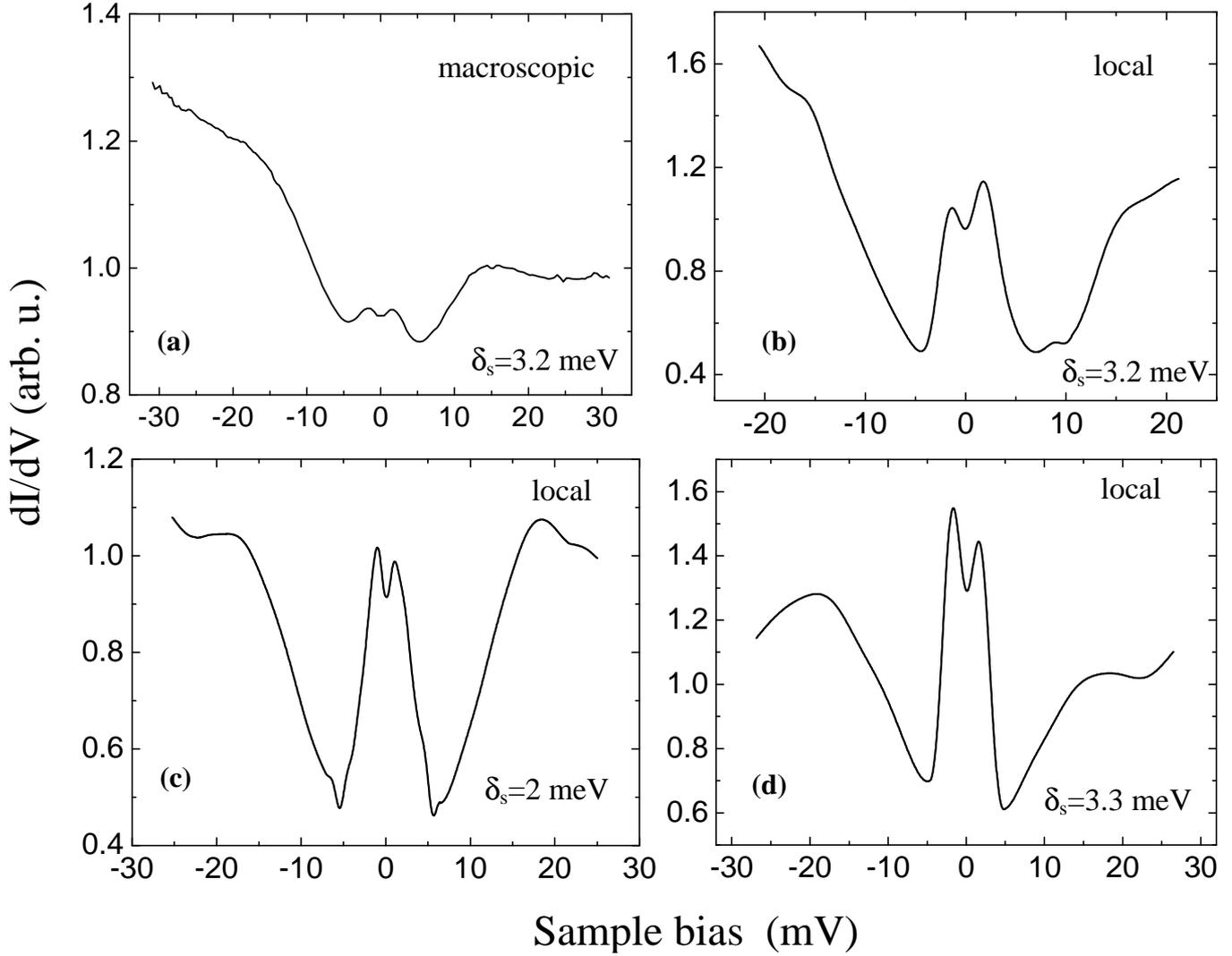

FIG 6. Splitting of the ZBCP in 5% Ca-YBCO film. (a) Macroscopic (indium pressed) tunnel junction with a 3.2 meV split peak inside a gap. (b-d) Local (STM) spectra measured on nominally identical film, with $\delta_s$ denoted in the images. Spectrum (b) has similar split and structure as the macroscopic data of (a). (c) and (d) depict smallest and largest observed $\delta_s$ values, respectively.

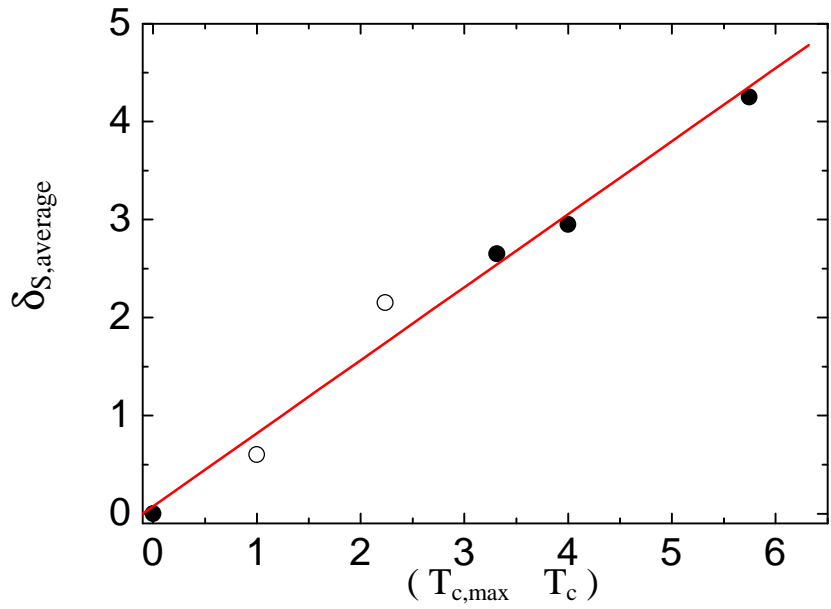

FIG 7. Doping dependence (proportional to ($T_{c,max}$ - $T_C$)) of the average ZBCP split size, as measured by STM. The straight line is a linear fit to the data. $T_c$ is taken at the point of zero resistance, and $T_{c,max}$ is the value for the optimally doped samples. Full circles: data measured for the (110) films discussed in this paper, empty circles: data measured for (001) YBCO films, discussed in Ref. 11.